\input epsf
\documentstyle[prd,floats,aps]{revtex}

\def\ut#1{\rlap{\lower1ex\hbox{$\sim$}}#1{}}

\begin{document}

\title{Towards a loop representation for quantum canonical supergravity}

\author{Daniel Armand-Ugon, Rodolfo Gambini}
\address{Instituto de F\'isica, Facultad de Ciencias,
Tristan Narvaja 1674, Montevideo, Uruguay}

\author{Octavio Obreg\'on}
\address{Instituto de F\'{\i}sica, Universidad de Guanajuato,
PO Box E-143, C. P. 37150 \\ Le\'on, Guanajuato,
M\'exico\\
and\\
Departamento de F\'{\i}sica,
Universidad Aut\'onoma Metropolitana-Iztapalapa,\\
PO Box 55-534, M\'exico, DF,
CP 09340, Mexico}

\author{Jorge Pullin }
\address{Center for Gravitational Physics and Geometry,
Department of Physics, 104 Davey Lab,\\
The Pennsylvania State University,
University Park, PA 16802. }

\maketitle
\begin{abstract}

We study several aspects of the canonical quantization of supergravity
in terms of the Asthekar variables.  We cast the theory in terms of a
$GSU(2)$ connection and we introduce a loop
representation.  The solution space is similar to
the loop representation of ordinary gravity, the main difference being
the form of the Mandelstam identities.  Physical states are in general
given by knot invariants that are compatible with the $GSU(2)$
Mandelstam identities.  There is an explicit solution to all the
quantum constraint equations connected with the Chern-Simons form,
which coincides exactly with the
Dubrovnik version of the Kauffman Polynomial. This provides for the first
time the possibility of finding explicit analytic expressions
for the coefficients of that knot polynomial.

\end{abstract}
\vspace{-12cm}
\begin{flushright}
\baselineskip=15pt
CGPG-95/8-3  \\
gr-qc/9508036\\
Revised November 21 1995
\end{flushright}
\vspace{11.3cm}

\section{Introduction}

Supersymmetry \cite{BaWe} provides a unified framework for the
discussion of bosons and fermions and has been proposed as a mechanism
to improve the behavior of quantum field theory divergences.  It may
offer the only viable alternative at present for grand unified models
\cite{Ca}.
Because it allows to write the Hamiltonian in terms of its ``square
roots'' it may provide a new perspective on the constraint equations
\cite{DeHaOb,GrCs}.

Supergravity was cast in a canonical fashion several years ago by
Deser, Kay and Stelle \cite{DeKaSt} , Pilati \cite{Pi} and D'Eath
\cite{De}.  Teitelboim \cite{TeTa} pointed out that it
provided a square root of the Wheeler-DeWitt equation. Jacobson
\cite{Ja} showed that new canonical variables similar to the ones
introduced by Ashtekar in general relativity could be used in
supergravity.  Here we will follow this latter approach.

In this paper we will show that several ideas that have appeared in the
context of the loop quantization of general relativity in terms of the
Ashtekar new variables \cite{As} find a natural counterpart in supergravity.

There are already partial results concerning the use of new variables
and loops in the context of connection representations for
supergravity.  F\"ul\"op \cite{Fu} first noticed that the theory could
be cast in terms of an $GSU(2)$ connection and discussed the
Chern-Simons form as a possible state. This state had already been
considered in a semi-classical context by Sano and Shiraishi
\cite{SaSh}.  Matschull \cite{Ma} noted that bosonic Wilson loops were
spurious solutions to all the constraints of supergravity.

In this paper we will use $GSU(2)$ Wilson loops with nontrivial
fermionic content as an overcomplete basis in terms of which we can
expand quantum state.  We
discuss the exponential of the $GSU(2)$ Chern-Simons form as a
diffeomorphism invariant solution to all the constraints of supergravity
with a cosmological constant with non-trivial fermionic content.  We
will also build a loop representation for the theory and discuss the
kinematical state space.  We will see that the loop counterpart of the
Chern-Simons solution is associated with the Dubrovnik version of
the Kauffman polynomial, which is therefore
compatible with the $GSU(2)$ Mandelstam identities.

\section{Quantum supergravity in terms of a $GSU(2)$ connection}

When cast in a canonical form in terms of the Ashtekar new variables
\cite{As}, supergravity is described by variables
$\tilde{E}^a_i,A_a^i,\psi_a^A,\tilde{\pi}^a_A$ where lowercase indices from
the beginning of the alphabet are spatial tensor indices, lowercase
indices from the middle of the alphabet are $SU(2)$ indices and
uppercase indices are spinor indices. The $E$'s are densitized triads,
the $A$'s are the Sen connection, the $\psi$'s are Grassman-valued
Rarita-Schwinger fields and the $\pi$'s are their canonically
conjugate momenta. We refer the reader to the paper by Jacobson
\cite{Ja} for details. We will follow the notation of F\"ul\"op
\cite{Fu}.

The canonical framework (with a cosmological constant) has the
following constraints,

\begin{eqnarray}
{\cal G}_i &=&
D_a \tilde{E}^a_i + {i\over \sqrt{2}} \pi^a_A \psi_{aB}
\sigma^{AB}_i=0
\label{gauss}\\
{\cal S}_A &=& D_a \pi^a_A -i \alpha \tilde{E}^a_i \sigma_{iA}^B \psi_{aB}\\
\bar{\cal S}^A &=& \epsilon^{ijk} \tilde{E}^a_i \tilde{E}^b_j
{\sigma_k}^A_B (-4 i D_{[a} \psi_{b]}^B+\sqrt{2}\bar{\alpha}
\ut{\epsilon}_{abc}
\pi^{cB}) =0
\end{eqnarray}
where $\sigma^{AB}_i$ are usual Pauli matrices and $\alpha$ and
$\bar{\alpha}$ are such that the cosmological constant is given by
$\Lambda = -\alpha \bar{\alpha}$.

The usual Hamiltonian and diffeomorphism constraints of general
relativity (with the corresponding fermionic extra terms) are obtained
by taking Poisson brackets of the ${\cal S}$ and $\bar{\cal S}$
constraints.

One could construct a quantum representation by considering wavefunctions
$\Psi(A,\psi)$ and promoting the above constraints to wave equations.
Matschull observed that if one does so and considers purely bosonic
wavefunctions consisting of the Wilson loops built with the bosonic
connection $A$ along smooth non-intersecting loops, all the contraints
are solved. This presented a puzzle, since such wavefunctions are
evidently not diffeomorphism invariant. How could they therefore
provide a solution to the diffeomorphism constraint? The answer is
given by the fact that the commutator of ${\cal S}$ with $\bar{\cal
S}$ give expressions for the usual Hamiltonian and diffeomorphism
constraint multiplied by the determinant of the spatial metric. Since
Wilson loops based on smooth loops are annihilated by the determinant
of the metric, they automatically solved the resulting constraints.

Here we will proceed in a different way. F\"ul\"op noticed that if one
considers the Gauss law and the right supersymmetry generator, they
form under Poisson brackets a graded Lie algebra associated with the
$GSU(2)$ group \cite{PaRi},

\begin{eqnarray}\label{4}
\{{\cal G}_i,{\cal G}_j\} &=& \sqrt{2} \epsilon_{ijk} {\cal G}_k\\
\{{\cal G}_j, {\cal S}_A\} &=&-
{i\over \sqrt{2}} (\sigma_j)^B{}_A
{\cal S}_B\\\label{6}
\{{\cal S}_A,{\cal S}_B\} &=& -i \alpha (\sigma^j)_{AB} {\cal G}_j.
\end{eqnarray}

In view of this, one can define new $GSU(2)$ variables $\tilde{\bf
E}^a_I$, ${\bf A}_a^J$, ${\bf F}_{ab}^K$,
\begin{eqnarray}
\tilde{\bf E}^a_I &=& ( \tilde{E}^a_i,\tilde{\pi}^a_A)\\
{\bf A}_a^I &=& (A_a^i,\psi_a^A)\\
{\bf F}_{ab}^I &=& (F_{ab}^i,2 D_{[a} \psi_{b]}^A)
\end{eqnarray}

The uppercase indices from the middle of the alphabet range from 1 to
5 and refer to a basis for the fundamental  representation of
$GSU(2)$, given by matrices ${\bf G}_I$,

\begin{equation}
{\bf G}_1 = -{i\over \sqrt{2}}
\left(\begin{array}{ccc} 0&1&0\\1&0&0\\0&0&0\end{array}\right),\quad
{\bf G}_2 = -{i\over \sqrt{2}}
\left(\begin{array}{ccc} 0&-i&0\\i&0&0\\0&0&0\end{array}\right),\quad
{\bf G}_3 = -{i\over \sqrt{2}}
\left(\begin{array}{ccc} 1&0&0\\0&-1&0\\0&0&0\end{array}\right),
\end{equation}

\begin{equation}
{\bf G}_4 =
\sqrt{\alpha\over \sqrt{2}}
\left(\begin{array}{ccc} 0&0&1
\\0&0&0\\0&1&0\end{array}\right),\quad
{\bf G}_5 =
\sqrt{\alpha\over \sqrt{2}}
\left(\begin{array}{ccc} 0&0&0\\0&0&1\\-1
&0&0\end{array}\right),\quad
{\bf e} =
\left(\begin{array}{ccc} 1&0&0\\0&1&0\\0&0&0\end{array}\right),
\end{equation}
where we have included the matrix ${\bf e}$ such that
$({\bf G}_i)^2=-{\bf e}/2$ for future purposes. The matrices satisfy
the commutation relations of the algebra (\ref{4}-\ref{6}).

The $GSU(2)$ group has a Killing-Cartan metric associated with
the orthosymplectic form
\begin{equation}
g_{IJ} \lambda^I \lambda^J =
\delta_{ij} \lambda^i \lambda^j -\sqrt{2} \alpha
\eta_{AB} \lambda^A \lambda^B
\end{equation}
where the $\lambda$'s are parameters in
the group with bosonic components $\lambda_i$ and anticommuting
components $\lambda_{A}$ and $\eta_{AB}$ is the antisymmetric symbol in
two dimensions. One has the usual normalization of the generators of
$GSU(2)$ in the case $\alpha =\sqrt{2}$, the other cases correspond to
a constant rescaling of the generators.

A remarkable fact is that
in terms of the $GSU(2)$ connection one
can introduce a covariant derivative ${\bf D}_a$ such that the Gauss
law and the right supersymmetry constraint can be written as a single
$GSU(2)$ Gauss law \cite{Fu},

\begin{equation}
{\bf D}_a \tilde{\bf E}^a_I =0.
\end{equation}

The left supersymmetry constraint can be written in terms of these
variables but it is not invariant under $GSU(2)$ transformations. This
is reasonable since it has a nontrivial Poisson bracket with the right
generator, giving rise to the usual Hamiltonian and diffeomorphism
constraints. In terms of these variables the left supesymmetry
constraint can be written as,
\begin{equation}
\bar{\cal S}_A = {\bf g}_A{}^{IJ}{}_K (\tilde{\bf E}^a_I \tilde{\bf E}^b_J
{\bf F}_{ab}^K +i \Lambda
\ut{\epsilon}_{abc}
\tilde{\bf E}^a_I \tilde{\bf E}^b_J \tilde{\bf E}^{cK})
\end{equation}
where
\begin{equation}
{\bf g}_A{}^{ij}{}_B \equiv -2 i \epsilon^{ij}_k \sigma^k{}_{AB}
\end{equation}
and all other components vanish. The definition of $\bf g$ is
obviously non $GSU(2)$ covariant.

The point of writing the equations in terms of these variables is that
it allows to find in a straightforward manner several solutions to
all the quantum constraint equations. Some of these solutions have a quite
nontrivial form when decomposed in terms of the original $SU(2)$ variables.
In order to see this let us consider a quantum representation in which
wavefunctions are functionals of the $GSU(2)$ connection $\Psi({\bf A})$.
The operator $\hat{\tilde{\bf E}}^a_I$ is represented as a functional
derivative whereas the operator $\hat{\bf A}_a^I$ is multiplicative.

It is immediate to find  solutions to the $GSU(2)$ Gauss law. Any
$GSU(2)$ invariant expression will do. In particular one can
consider  Wilson loops constructed with the $GSU(2)$
connection,
\begin{equation}
W_\gamma ={\rm STr P}\exp\left(\oint_\gamma dy^a {\bf A}_a\right)
\end{equation}
where $\gamma$ is a loop on the spatial manifold. The trace taken in
the above expression is the supertrace \cite{DeWi}, which for any $GSU(2)$
matrix $A$ is given by ${\rm STr}(A)=A_{11}+A_{22}-A_{33}$.

Since it is $GSU(2)$ invariant, the Wilson loop is annihilated by the Gauss
law. These states are therefore invariant under
right supersymmetry transformations and triad
rotations. If one studies the action of the left supersymmetry
constraint on them it is immediate to notice that if one
considers loops $\gamma$ that are smooth (they do not have kinks or
intersections), the constraint annihilates these states. The
reasons are the same as the ones that made the usual Ashtekar Hamiltonian
constraint annihilate a Wilson loop based on a smooth loop \cite{JaSm}.
The states
have a quite nontrivial fermionic content, as we will discuss in the
next section (they correspond to an infinite superposition of terms
built with holonomies with arbitrarily high number of fermionic
insertions along the loop, if $\alpha\neq0$). Notice that these states
are different from those of Matschull \cite{Ma} which were purely
bosonic. They reduce to them in the case $\alpha=0$. They also share
with those states the  pathology that Matschull pointed out:
although they solve all the
constraints they are not diffeomorphism invariant. We will however
find that
related states are useful for the construction of a loop
representation in the next section.

It is possible to find another exact state that solves all the
constraint equations. This state is genuinely diffeomorphism invariant
and is associated with a non-degenerate spatial metric (with non-vanishing
determinant). If one
considers the state built by taking the exponential of the
Chern-Simons form of the $GSU(2)$ connection,

\begin{equation}
\Psi_\Lambda({\bf A}) = \exp({i\over 2 \Lambda}
\int d^3x {\rm STr}({\bf A}_a \partial_b {\bf A}_c +
{\sqrt{2} i \over 3}{\bf A}_a {\bf A}_b {\bf A}_c) \tilde{\epsilon}^{abc}
\end{equation}
it is immediate to see that it satisfies all the supersymmetry
constraints.  It is annihilated by the $GSU(2)$ gauss law since it is
$GSU(2)$ invariant.  It is annihilated by the left supersymmetry
constraint for the same reason the corresponding state was annhilated
by the Hamiltonian constraint of quantum gravity with a cosmological
constant \cite{BrGaPu}: the state has the property $\hat{\tilde{\bf
E}}^a_I \Psi(A) ={i \over 2\Lambda} \tilde{\epsilon}^{abc}{\bf F}_{bcI}
\Psi(A)$ and therefore the two terms in the left supersymmetry
constraint are identical and of opposite sign, cancelling
each other.  Being the integral of a three form, the state is naturally
diffeomorphism invariant.  Some concerns may be raised about factor
ordering since in the ordering with the triads to the left (where the
state is a solution) the constraint that usually corresponds to
diffeomorphisms formally does not generate that symmetry.  As was
discussed in \cite{BrGaPu}
using a symmetric regularization for the constraint
it actually generates diffeomorphisms and annihilates the state and
similar considerations apply here.  This state was first introduced by
Kodama \cite{Ko} for usual gravity and was discussed in a semiclassical
supergravity context by Sano and Shiraishi \cite{SaSh}.
F\"ul\"op \cite{Fu} noticed that it
was an exact solution of the Wheeler-DeWitt equation of full supergravity.
All these discussions were in terms of the $SU(2)$
variables.  By writing the state and the constraints in terms of the
$GSU(2)$ variables we notice here that it is very easy to see why it
is annihilated by the constraints.  It will also be the key to finding
a counterpart of this state in the loop representation.

The Chern-Simons form has been observed to be a solution of all the
constraints of Maxwell theory, Yang-Mills \cite{Ja83}, Einstein
\cite{BrGaPu}, Einstein-Maxwell \cite{GaPu} and also supergravity. It
is remarkable that so different theories have a similar state in
common.

\section{Loop representation}

The usual starting
step to construct a loop representation \cite{GaTr,RoSm90} for
a theory based on a connection is to expand the states of the theory
in terms of a basis of functions given by the Wilson loops constructed
with the connection. Such basis
is gauge invariant and the resulting representation (the loop
representation) is therefore well suited for the description of gauge
invariant operators. Crucial to this construction is the knowledge
that every gauge invariant quantity can be expanded in terms of Wilson
loops. For the bosonic case this last statement is the content of a
theorem by Giles \cite{Gi}. For the supersymmetric case we do not know if a
similar theorem holds. Even if a theorem like the one above holds, in
general one can only expand wavefunctions in terms of products of
Wilson loops. In many theories one can re-express an arbitrary product
of Wilson loops as a linear combination of products of a fixed minimal
number of Wilson loops. The resulting loop representation is based on
wavefunctions depending on multiloops of at most that minimal number,
through the relation,

\begin{equation}
\Psi(\gamma_1,\gamma_2,\ldots,\gamma_n) =
\int dA W_{\gamma_1}[A] W_{\gamma_2}[A]\cdots W_{\gamma_n}[A] \Psi[A]
\end{equation}
where $n$ is the minimal number of Wilson loops in terms of which one
can express an arbitrary product of Wilson loops.

To address the two issues mentioned above, namely if Wilson loops are
enough to represent any state and which is the minimal number of Wilson
loops needed, let us discuss some properties of the supesymmetric Wilson
loops.

Because they are traces of group elements, the Wilson loops satisfy
certain identities called Mandelstam identities \cite{GaTr86}
which reflect particular properties of the group in question. For the
group we are considering they are rather nontrivial, so we start with a
discussion for the case with $\Lambda=0$. They are given by,
\begin{eqnarray}
W_{\gamma_1\circ \gamma_2} &=& W_{\gamma_2\circ \gamma_1},\\
W_{\gamma_1} W_{\gamma_2} &=&
W_{\gamma_1\circ\gamma_2} +
W_{\gamma_1\circ\gamma_2^{-1}} -
W_{\gamma_1}-W_{\gamma_2}+1.\label{mandelfund}
\end{eqnarray}

These identities allow to express any product of Wilson loops as a
linear combination of Wilson loops. They can be combined in many
nontrivial ways\footnote{For the usual loop representation of quantum
gravity, which is based on an $SU(2)$ group, the identity
(\ref{mandelfund}) reads $
W_{\gamma_1} W_{\gamma_1} =
W_{\gamma_1\circ\gamma_2} +
W_{\gamma_1\circ\gamma_2^{-1}}$.}. For instance, it follows from
(\ref{mandelfund}) taking $\gamma_1$ to a point that,

\begin{equation}
W(\gamma) = W(\gamma^{-1}).
\end{equation}

In order to derive these identities one needs to recall that a generic
element of the group in the case $\Lambda=0$
is written as ${\bf 1} +\phi_0 {\bf e} +
\phi_I {\bf G}^I$ with $(1+\phi_0)^2+\sum_{I=1}^3 \phi_I^2 =1$.

For the case $\Lambda\neq 0$ it is harder to find the identities.  We
have only suceeded in finding part of them.  In order to do this we make
use of the techniques introduced in \cite{UrWaZe} that related the
Mandelstam identities with the Cayley-Hamilton theorem.  This theorem
states that any matrix is a root of its characteristic polynomial.  For
supermatrices a generalization of this statement has been discussed in
reference \cite{UrMo}.  There it was established that a supermatrix is a
root of a polynomial associated with the characteristic polynomial,
which acts as a generalization of the Cayley-Hamilton polynomial to the
case of supermatrices.  For a generic matrix of the orthosymplectic
group (which includes as a subgroup $GSU(2)$) the Cayley-Hamilton
equation is \cite{UrWaZe},
\begin{equation}\label{21}
{\bf H}^3(\gamma) -({\rm STr}({\bf H}(\gamma))+2)
({\bf H}^2(\gamma)-{\bf H}(\gamma))-1 =0,
\end{equation}

If one now
multiplies it by ${\bf H}(\eta) {\bf H}(\gamma)^{-1}$ and takes the
trace one gets,
\begin{eqnarray}
&&{\rm STr}({\bf H}(\eta){\bf H}(\gamma)^2))-
({\rm STr}({\bf H}(\gamma))+2)
{\rm STr}({\bf H}(\eta){\bf H}(\gamma))+\nonumber\\
&&\quad({\rm STr}({\bf H}(\gamma))+2)
({\rm STr}({\bf H}(\eta)))
-{\rm STr}({\bf H}(\eta){\bf H}(\gamma)^{-1})=0.\label{mandid}
\end{eqnarray}

This is a ``Mandelstam identity'' in the sense that it is satisfied by
the traces of the holonomy we are considering. However one is usually
interested in identities that allow to reduce the number of traces.
{}From this point of view, the above identity does not help. One
can derive another identity that allows to reduce the number of traces
in a product by considering the generalization of formula (21) to the
case of generic $(2,1)$ matrices\footnote{An
arbitrary $3\times 3$ matrix in which the $13$, $23$, $31$ and $32$
components are Grassmanian.} \cite{BeUr}.
One then considers a $(2,1)$
matrix given by a sum of $GSU(2)$ matrices and inserts it in the
identity.
The resulting expression relates products of traces of the $GSU(2)$
matrices. and allows to expand an arbitrary product of Wilson loops as a
linear combination of products of three Wilson
loops. When we build the loop representation it will therefore be
based on wavefunctions depending at most on three loops. There is yet
another Mandelstam identity that has to be considered, that which
reflects the fact that the matrices are of unit determinant. We will not
consider it here.

The use of Cayley-Hamilton techniques allows us to state that in
general, for any theory based on a group or supersymmetric group the
Cayley-Hamilton theorem can be used to prove that the loop
representation will involve wavefunctions of a finite number of loops.

Let us address the question of up to what extent one can use
combinations of these Wilson loops to represent any gauge invariant
state.  As we mentioned before, in the bosonic case this was ensured by
Giles' theorem.  Here we do not have a similar theorem.  However it is
easy to see that it is likely to be a problem.  Consider the case
$\Lambda=0$.  As we saw, the supersymmetric Wilson loops reduce to a
purely bosonic expression coinciding with the Wilson loop constructed
with the bosonic $SU(2)$ connection.  Therefore they clearly fail to
capture any fermionic information.  For the $\Lambda\neq 0$ case we do
not know what the situation exactly is.  It should be noted that the
$\Lambda=0$ case is somewhat pathological from the point of view of the
$GSU(2)$ symmetry we are exploiting since the invariant orthosymplectic
form collapses to the usual Euclidean form.  We will proceed to build a
pure loop representation to highlight other aspects of the construction
but it should be forewarned that it is possible that the resulting
quantum representation only captures part of the information present in
the theory.

Apart from the above mentioned peculiarities of the supersymmetric case,
there is yet another difference with the usual loop representation of a
bosonic theory. The $GSU(2)$ Wilson
loops naturally implement the symmetry of the theory under triad
rotations and right supersymmetry transformations. If we construct a
loop representation using them as a basis for states it will be
difficult to write an expression in such a representation for the left
supersymmetry generator $\bar{\cal S}$, which is not invariant under
$GSU(2)$ rotations. It is like trying to represent in the usual loop
representation for a Yang-Mills theory
a non-gauge invariant quantity. A similar
situation arises in gravity when one considers the diffeomorphism
constraint. The space of solutions to this constraint is given by
functions of knots. In this space we cannot represent the Hamiltonian
constraint, since it is not diffeomorphism invariant. This forces us to
define the Hamiltonian constraint on functions of loops rather than on
functions of knots, in spite of the fact that the space of solutions to
the constraint that are physical will be given by functions of knots.

Here we will proceed in a similar way. We will build a representation in
terms of loops and an extra structure, in order to be able to represent
the left supersymmetry constraint. The solutions to the constraint
however, will be given by pure functions of loops.

In order to accomplish this let us
examine in detail the expression for the $GSU(2)$
Wilson loop we introduced above. If one expands out the expression for
the path ordered exponential one has that,

\begin{equation}
W_\gamma({\bf A}) = 1 +\sum_{n=1}^\infty \oint_\gamma dy_1^{a_1}
\cdots \int_o^{y_{n-1}} dy_n^{a_n} {\rm Tr}({\bf A}_{a_1}(y_1)\ldots
{\bf A}_{a_n}(y_n)).
\end{equation}

To further discuss this expression we introduce the following
notation: the non-boldface matrix $A$ will represent the combination
${\bf A}_a^i
G_i$, $i=1\ldots 3$ and the matrix $\psi_a$ will represent the
combination ${\bf A}_a^B G_B$, $i=4,5$. Both are $GSU(2)$
matrices. In terms of these quantities the above expression can be
written as,
\begin{eqnarray}
W_\gamma({\bf A}) &=& 1+\sum_{n=1}^\infty
\oint_\gamma dy_1^{a_1}
\cdots \int_o^{y_{n-1}} dy^{a_n}_n {\rm STr}(A_{a_1}(y_1)\ldots
\ldots A_{a_n}(y_n))\nonumber\\
&&+\sum_{n=2}^\infty \sum_{i=1}^n
\oint_\gamma dy_1^{a_1}
\cdots \int_o^{y_{n-1}} dy^{a_n}_n{\rm STr}(A_{a_1}(y_1)\ldots
A_{a_{i-1}}(y_{i-1}) \psi_{a_i}(y_i) \ldots A_{a_n}(y_n))\nonumber\\
&&+\sum_{n=2}^\infty \sum_{j=2}^n \sum_{i=1}^{j-1}
\oint_\gamma dy_1^{a_1}
\cdots \int_o^{y_{n-1}}  dy^{a_n}_n\\&&\quad\times{\rm
STr}(A_{a_1}(y_1)\ldots
A_{a_{i-1}}(y_{i-1}) \psi_{a_i}(y_i) \ldots
A_{a_{j-1}}(y_{j-1}) \psi_{a_j}(y_j) \ldots A_{a_n}(y_n)). \nonumber
\end{eqnarray}

The above expression can be rearranged as,

\begin{eqnarray}
W_\gamma({\bf A}) &=& \sum_{n=0}^\infty
GT^n_{\gamma}({\bf A})\label{wila}\\
GT^n_{\gamma}({\bf A})&\equiv& \oint_\gamma dy_1^{a_1}
\cdots \int_o^{y_{n-1}} dy^{a_n}_n
{\rm Tr}(\psi_{a_1}(y_1) U(\gamma_{y_1}^{y_2})
\psi_{a_2}(y_2) \ldots U(\gamma_{y_{n}}^{y_1})).
\end{eqnarray}

Each of the terms in the summation in (\ref{wila}) is $SU(2)$ invariant.
In order to see this notice that the Gauss law (\ref{gauss}) acts
homogeneously in $\psi$ and therefore does not mix the different terms
in (\ref{wila}).  Since the sum is invariant, each term has to be
invariant as well.  This can be explicitly checked and one notices that
it occurs in a rather nontrivial fashion, using explicitly the fact that
the $\psi$'s are Grassman-valued.  These invariants can be viewed as
holonomies that one breaks at certain points and inserts a $\psi$ and
then integrates the point along the loop.  These invariants are quite
different than the ones one usually considers when building loop
representations for theories coupled to fermions, where the fermions can
only appear at the ends of open paths and is directly connected with the
$GSU(2)$ invariance that is characteristic of this theory.

One can build a loop representation based on the invariants
introduced above either through a transform or through the
introduction of further invariants that involve the momenta ${\bf E}$.
We will choose the first approach for brevity. Given a wavefunction in
the connection representation $\Psi({\bf A})$ one constructs a
wavefunction in the loop representation by,

\begin{equation}
\Psi(\epsilon,\gamma) = \sum_{i=1}^\infty \int d {\bf A} \epsilon^n
GT^n_{\gamma}({\bf A}) \Psi({\bf A})
\end{equation}

Wavefunctions in the loop representation are characterized by a loop
and an real parameter $\epsilon$.

The $GT^n_{\gamma}$'s satisfy a series of identities similar to the
Mandelstam identities for the Wilson loops we introduced at the
beginning of this section, we will not discuss them here since we will
not need them for the issues addressed in this paper.

In this representation the left supersymmetry constraint cannot be
directly represented since it is not $SU(2)$ invariant. However one
can build very easily $SU(2)$ invariant expressions that are equivalent
to the left supersymmetry constraint, for instance, by contracting it
with the spin $3/2$ field.

The fact that the representation is not completely cast in terms of
loops appears to detract from the geometric nature of the usual loop
representation. However, as we pointed out, one needs the extra
parameter to represent non $GSU(2)$ invariant quantities and states. The
solutions to the constraint are $GSU(2)$ invariants and therefore are
expressible purely in terms of loops. We will analyze in the following
section one such solution, the one that is obtained by transforming into
the loop representation the Chern-Simons state we discussed in the
previous section.

\section{The Chern-Simons state and the
Dubrovnik version of the Kauffman Polynomial
 as a solution to the
super-Wheeler--DeWitt equation}

Let us consider the expression in the loop representation of the
solution to all the constraint equations with a cosmological constant
given by the exponential of the super Chern-Simons form,

\begin{equation}
\Psi(\gamma_1,\ldots,\gamma_n) = \int dA e^{{i\over 2 \Lambda} S_{CS}}
W_{\gamma_1}[A] W_{\gamma_2}[A]\cdots W_{\gamma_n}[A]
\end{equation}
where $S_{CS}$ is the super Chern-Simons form we introduced above. Since
it is a $GSU(2)$ invariant state we only need to introduce Wilson loops
in the transform.

Assuming that one is using a diffeomorphism invariant measure in the
transform, the resulting state in the loop representation has to be a
knot polynomial. We will show that it is the Kauffman Polynomial,
Dubrovnik version.

The derivation is analogous to the
one giving the Kauffman bracket as a state of ordinary bosonic gravity
in the loop representation and was discussed in reference
\cite{BrGaPu}.

We will first study the transform in  the case of a one component loop
\begin{equation}
\Psi(\gamma) = \int dA e^{{i\over 2 \Lambda} S_{CS}} W_\gamma[A].\label{30}
\end{equation}

\begin{figure}
\hspace{3.7cm}\epsfxsize=300pt \epsfbox{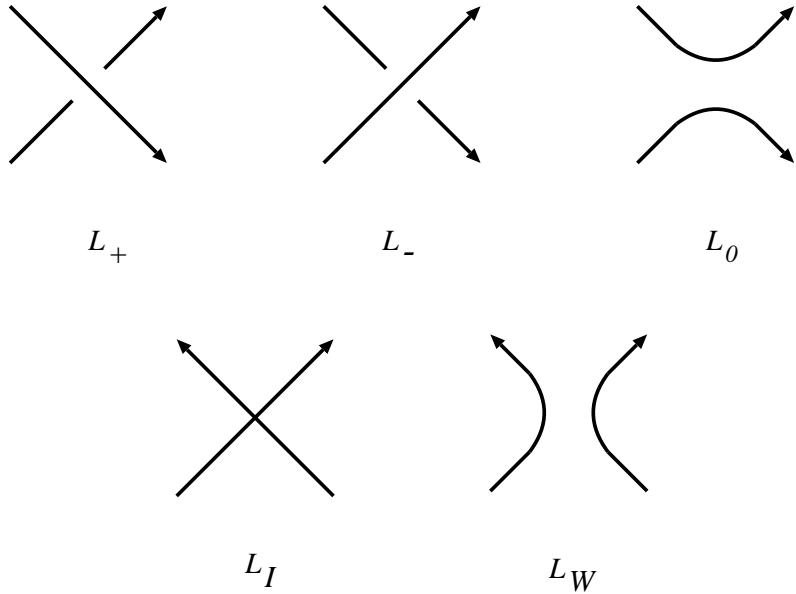}
\caption{The different crossings involved in the skein relations}
\label{skein}
\end{figure}

\begin{figure}
\hspace{2.7cm}\epsfxsize=400pt \epsfbox{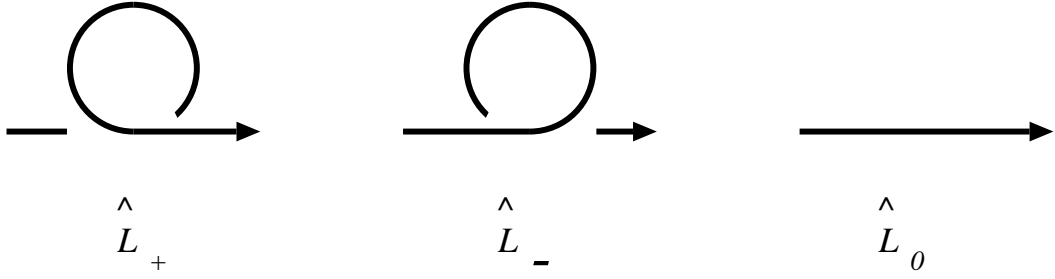}
\caption{Regular isotopy invariants are not invariant under
the addition and removal of a curl. This is determined by the skein relations
involving the elements shown}
\label{hats}
\end{figure}

In the case of bosonic gravity, the transform of the
Chern-Simons state is given by the HOMFLY polynomial.
The skein relations satisfied by a regular-isotopic\footnote{As in the
case of the bosonic transform of the Chern-Simons state the resulting
invariant is a function of framed links. In the bosonic case that is
the reason the result is the HOMFLY polynomial rather than the Jones
polynomial.} HOMFLY polynomial $F(z,t,\alpha)_\gamma$ are,

\begin{eqnarray}
F_U&=&1\\
F_{\hat{L}_+} &=& \alpha F_{\hat{L}_0}\\
F_{\hat{L}_-} &=& \alpha^{-1} F_{\hat{L}_0}\\
t F_{L_+} -t^{-1}  F_{L_-} &=& z F_{L_0}
\label{skein3}
\end{eqnarray}
where these relations are to be understood as follows. Given a knot,
pick a crossing in its planar diagram and replace it with either
$L_+$, $L_-$ or $L_0$ as depicted in figure \ref{skein} or figure
\ref{hats} for the hatted elements. Evaluate the polynomial on the
resulting links. The resulting polynomials are related by the above
expressions. The first equation is a normalization condition stating
that the polynomial evaluated on the unknot is equal to one.

We will show that the expression of the transform of the super Chern-Simons
state satisfies related but different skein relations. In order to do
this we will perform the following
computation first suggested by Smolin \cite{Sm} and Cotta-Ramusino et
al \cite{CoGuMaMi}. Starting from the expression of the transform
evaluated at an intersection we will append to it an infinitesimal
loop in such a way as to turn the intersection into an over
crossing. We will then repeat the same procedure to turn it into an
under crossing. We see that the difference of the resulting
expressions is related to the value of the transform evaluated at
$L_0$ through a skein relation of the same nature of  (\ref{skein3}), but
for a different polynomial.

\begin{figure}
\hspace{5.7cm}\epsfxsize=100pt \epsfbox{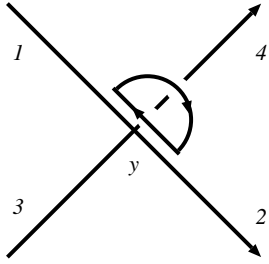}
\caption{The addition of a small loop at an intersection}
\label{inter}
\end{figure}

The variation of the loop transform of the state when a small loop of
element of area $\sigma^{ab}$ is appended to a single-component loop
$\gamma$ at an intersection as shown in figure \ref{inter} is given
by,

\begin{equation}
\sigma^{ab} \Delta_{ab}(y) \Psi[\gamma] = -2 \Lambda i (-1)^P
\int dA \sigma^{ab}
\epsilon_{dab} {\rm STr}[{\bf G}_I {\bf H}_{23}(\gamma_y^y)
{\bf H}_{41}(\gamma_y^y)] {\delta \over \delta
A_d^I(y)} g^{IJ} {\rm exp}({\textstyle {i\over 2\Lambda}} S_{CS})
\end{equation}
where $\Delta_{ab}$ is the loop derivative \cite{GaPubook}, $G_I$ is
one of the generators of $GSU(2)$ and we have used $\Delta_{ab}(x)
{\rm STr}[{\bf H}(\gamma)] = {\bf F}_{ab}^I(x)
{\rm STr}[{\bf G}_I {\bf H}(\gamma_x^x)]$ and $\gamma_x^x$
is the loop with origin at the point $x$. The labels on the holonomy
indicate the connectivity at the intersection, for instance $H_{23}$
is the loop that begins at $2$ and ends at $3$ as indicated in the
figure. We have also used the fundamental property of the Chern-Simons
state that we discussed in section 3,

\begin{equation}
{\delta \over \delta {\bf A}_a^I} \Psi_{\Lambda}[A] =
{i \over 2 \Lambda}
\tilde{\epsilon}^{abc} {{\bf F}_{bc}}_I \Psi_{\Lambda}[A]
\end{equation}
to convert the $F_{ab}$ factor due to the loop derivative into a
functional derivative acting on the exponential of the Chern-Simons
action. It should be recalled that indices like $I$ are raised and lowered
with the orthosymplectic metric $g^{IJ}$. The factor $(-1)^P$ is
introduced to take care of the flips in sign that take place when
$F_{ab}$ is moved from the left of the supertrace to the right of it
before converting it into a functional derivative and
is determined by the odd/even nature of each of the components of
$F_{ab}$ and the supertrace.

Integrating by parts and
choosing the element of area $\sigma^{ab}$ parallel to segment 1-2 so that the
contribution of the functional derivative corresponding to the action on the
segment 1-2 vanishes (since the volume element is zero) we get,

\begin{eqnarray}
\sigma^{ab} \Delta_{ab} \Psi[\gamma] &=&
2 i \Lambda  \int dA
\sigma^{ab} \epsilon_{abc} \int dv^c \delta(y-v) \times \nonumber\\
&& {\rm STr}[ {\bf G}_I {\bf H}_{23}(\gamma_y^y)
{\bf G}_J {\bf H}_{41}(\gamma_y^y) ] g^{IJ}
{\rm exp} ({\textstyle {i \over 2 \Lambda}} S_{CS})(-1)^{JH},
\end{eqnarray}
and in the integration by parts the $(-1)^P$ factor is cancelled but a
factor $(-1)^{JH}$ is introduced when the functional derivative is
``introduced'' in the supertrace at the end of the holonomy going from
$2$ to $3$ and is determined by the odd/even nature of the connection in
the functional derivative and the components of the
holonomy going from $2$ to $3$. As is usual in these kinds of
variational derivations \cite{Br}, a regularization of the volume element
determined by the element of area of the loop derivative and the tangent
to the loop is
needed, we take it in such a way that the volume is normalized to be
$\pm1$ depending on the orientation.

We now make  use of the Fierz identity for $GSU(2)$,
\begin{eqnarray}
g^{IJ} {({\bf G}_I)^{\kappa}}_\beta {({\bf G}_J)^{\gamma}}_\delta &=&
{{\bf e}^\kappa}_\delta {{\bf e}^\gamma}_\beta-{1\over 2}
{{\bf e}^\kappa}_\beta {{\bf e}^\gamma}_\delta \nonumber\\
&&+{\textstyle {1\over 2}}(\delta^\kappa_1 \delta^3_\beta +\delta^\kappa_3
\delta^2_\beta)(\delta^\gamma_2 \delta^3_\delta-\delta^\gamma_3
\delta^1_\delta) \nonumber\\
&&-{\textstyle {1\over 2}}
(\delta^\kappa_2 \delta^3_\beta -\delta^\kappa_3
\delta^1_\beta)(\delta^\gamma_1 \delta^3_\delta+\delta^\gamma_3
\delta^2_\delta)
\end{eqnarray}
and taking into account the explicit general form of an element of
$GSU(2)$ \cite{To},
\begin{equation}
\left(\begin{array}{ccc}a (1-\alpha p q/2)&b (1-\alpha p
q/2)&\sqrt{\alpha} p\\c (1-\alpha p q/2)&d  (1-\alpha p
q/2)&\sqrt{\alpha} q\\\sqrt{\alpha} (-a q +c p)&
\sqrt{\alpha} (d p -b q)& (1+\alpha p q)\end{array}\right)\label{explicit}
\end{equation}
where $ad-bc=1$ and $p$ and $q$ are Grassmanian variables one finally gets,
\begin{eqnarray}
&&\sigma^{ab} \Delta_{ab} \Psi[\gamma] = \label{potskein}\\
&=&  i \Lambda \int dA \;
\sigma^{ab} \epsilon_{abc} \int dv^c \delta(y-v)
{\rm STr}[{\bf H}_{23}(\gamma_y^y)]
{\rm STr}[{\bf H}_{41}(\gamma_y^y)]
{\rm exp}({\textstyle {i \over 2\Lambda}}
S_{CS}) \nonumber\\
&+&  i \Lambda
\int dA \; \sigma^{ab} \epsilon_{abc} \int dv^c \delta(y-v)
{\rm STr}[{\bf H}_{23}(\gamma_y^y) {\bf H}^{-1}_{41}(\gamma_y^y)]
{\rm exp}( {\textstyle {i \over 2\Lambda}} S_{CS}).\nonumber
\end{eqnarray}

At this point we may proceed as in the bosonic case and reinterpret
relation (
\ref{potskein}) as a skein relation. The form of the expression we got
suggests that the skein relation is,
\begin{equation}
\Psi[L_\pm] -\Psi[L_I] =\pm i \Lambda (  \Psi[L_0] +
\Psi[L_W] ).\label{skeinbuena}
\end{equation}

There is a subtle difference, however, between expression (\ref{potskein})
and the skein relation (\ref{skeinbuena}).
The first involves a rerouting of a portion of
the loop. The implication of that rerouting for the connectivity of the
loop at the intersection (the only ingredient that participates in the skein
relation) is only determined given an initial connectivity of the loop at
the intersection. The above skein relation corresponds to starting with
a loop with a connectivity at the intersection corresponding to a single
loop. One could keep the same intersection and reconnect the strands in
such a way that the initial loop has two components. That requires a
separate derivation for the addition of an infinitesimal element of area.

\begin{figure}
\hspace{3.7cm}\epsfxsize=300pt \epsfbox{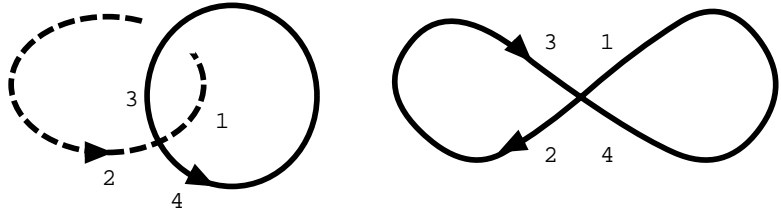}
\caption{The same ``straight-through'' crossing can correspond to a two loop
link or a single loop depending on the connectivity of the diagram.}
\label{multi}
\end{figure}

Let us therefore perform such a calculation. We consider a two
component loop with an intersection as shown in figure \ref{multi},
\begin{equation}
\Psi(\gamma_1,\gamma_2) = \int dA e^{{i\over 2 \Lambda} S_{CS}}
 W_{\gamma_1}[A]  W_{\gamma_2}[A].\label{42}
\end{equation}

The addition of a small loop at the intersection gives, through a
calculation very similar to the one performed in the case of a single
component loop,
\begin{eqnarray}
\sigma^{ab} \Delta_{ab} \Psi(\gamma_1, \gamma_2) &=&
2 i \Lambda  \int dA
\sigma^{ab} \epsilon_{abc} \int dv^c \delta(y-v) \times \nonumber\\
&& {\rm STr}[ {\bf G}_I {\bf H}_{12}({\gamma_1}_y^y)] \times {\rm STr}[
{\bf G}_J {\bf H}_{43}({\gamma_2}_y^y) ] g^{IJ}
{\rm exp} ({\textstyle {i \over 2 \Lambda}} S_{CS})(-1)^{JH_{12}},
\end{eqnarray}
and the labels refer to the figure \ref{multi}.

Using the Fierz identity this expresion can be written as

\begin{eqnarray}
&&\sigma^{ab} \Delta_{ab} \Psi(\gamma_1,\gamma_2) = \label{44}\\
&=&  i \Lambda \int dA \;
\sigma^{ab} \epsilon_{abc} \int dv^c \delta(y-v)
{\rm STr}[{\bf H}_{12}({\gamma_1}_y^y)
          {\bf H}_{43}({\gamma_2}_y^y)]
{\rm exp}({\textstyle {i \over 2\Lambda}}
S_{CS}) \nonumber\\
&-&  i \Lambda
\int dA \; \sigma^{ab} \epsilon_{abc} \int dv^c \delta(y-v)
{\rm STr}[{\bf H}_{12}({\gamma_1}_y^y) {\bf H}^{-1}_{43}({\gamma_2}_y^y)]
{\rm exp}( {\textstyle {i \over 2\Lambda}} S_{CS}).\nonumber
\end{eqnarray}

These results can be interpreted as the following skein relation
for the intersection,
\begin{equation}
\Psi[L_\pm] -\Psi[L_I] =\pm i \Lambda ( \Psi[L_0]- \Psi[L_W])
\end{equation}
for the case of a two component link.

If we now replace $W(\gamma)$ by $\hat{W(\gamma)}= -W(\gamma)$ in the
above expression (\ref{30},\ref{potskein}) and in (\ref{42},\ref{44}), signs
change in such a way that both expressions can be associated with a
single skein relation,
\begin{equation}
\Psi[L_{+}] -\Psi[L_{-}] = -2 i \Lambda
( \Psi[L_0] - \Psi[L_W] ).
\end{equation}

\begin{figure}
\hspace{3.7cm}\epsfxsize=300pt \epsfbox{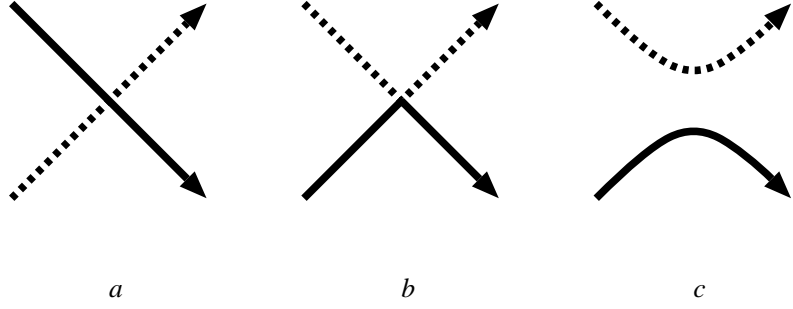}
\caption{Intersections without (a) and with  (b) kinks.
Under the present regularization intersections of type (b) are
equivalent to no intersection (c).}
\label{kinks}
\end{figure}

Some comments are in order. The study of the variational relation
obtained by adding an infinitesimal element of area to a crossing with
kinks proceeds along similar lines as the one we exhibited above. The
main difference is that the volume term has two contributions that add
up with opposite signs, corresponding to the addition of a small area
along the two different tangents that enter the kink. With a suitable
regularization, the whole contribution can be taken to be zero. This
implies that an intersection with a kink can be taken as equivalent to
no intersection at all, as shown in the figure \ref{kinks}, and this
is what allow us to replace the kinks by the $L_0$ and $L_W$
in equations (\ref{potskein},\ref{44}) to obtain the
skein relations.

To completely characterize the polynomial, we need the value of
the polynomial for the
unknot, which is -1, to be consistent with
the redefinition of the holonomy as minus the supertrace we introduced
above. Also, since the polynomial turns out to be
a regular isotopy invariant, we need to study the effect of
the addition of a ``curl'' at a point with no intersection. The
details are exactly the same as those in reference \cite{BrGaPu} so we
will omit them here. The only new ingredient needed is a contraction
of the Fierz identity,
\begin{equation}
g^{IJ} {({\bf G}_I)^\kappa}_\beta {({\bf G}_J)^\beta}_\gamma=
\delta^\kappa_\gamma,
\end{equation}
and one gets as result,
\begin{equation}
\Psi[\hat{L}_\pm] = (1 \pm 2 \Lambda i) \Psi[\hat{L}_0]
\end{equation}
where the meaning of the hatted elements is described in figure \ref{hats}.

Let us considered now the Dubrovnik version of the Kauffman Polynomial; it is
defined by the following relations \cite{onknots},

\begin{eqnarray}
\Psi[L_{+}] -\Psi[L_{-}] &=& z (  \Psi[L_0]- \Psi[L_W]) \\
\Psi[{\rm unknot}] &=& (a-a^{-1})/z  + 1 \\
\Psi[\hat{L}_{+}] &=& a \Psi[L_0]    \\
\Psi[\hat{L}_{-}] &=& a^{-1}\Psi[L_0]  \newline
\end{eqnarray}

One can check that this correponds exactly to the results we obtained
to first order provided $z= -2i\Lambda$ and $a = 1+2i\Lambda $ and $
\Psi[L_0] = -1 $ which is exactly the case.

We therefore see that the
Dubrovnik version of the Kauffman Polynomial bears the same relationship
with a $GSU(2)$ Chern--Simons theory as the Kauffman bracket has with
the usual $SU(2)$ theory. It is also
important to remember that we have shown that the Dubrovnik version of the
Kauffman polynomial is the transform into the loop representation of an
exact quantum state of supergravity. It therefore should follow (as it
happens in ordinary gravity) that the polynomial should be annihilated by
all the constraints of quantum supergravity.
It is also worthwhile pointing out  that one could explore
states without cosmological constant based on the ambient invariant
polynomial associated with the one we found, as happens in the bosonic
case \cite{essay}.

An interesting aspect is that Saleur and Zhang \cite{Sa,Zh} have
considered representations of the braid group associated with graded
groups and found associated knot polynomials. It would be interesting
to check if these have a relation with the Dubrovnik version of the
Kauffman polynomial, as the calculation we have performed suggests. It
would also be important to find ways to compute the skein relations we
found to first order in exact form. This could be accomplished via a
supersymmetric version of the Moore-Seiberg-Witten \cite{MoSe,Wi}
construction based on conformal field theory.  Finally, one could
evisage computing explicit expressions for each coefficient of the
Dubrovnik version of the Kauffman polynomial by perturbatively
evaluating the expectation value of a Wilson loop in Chern--Simons theory.
For the bosonic case this was first studied by Guadagnini, Martellini,
and Mintchev \cite{GuMaMi} and although more complex, the supersymmetric
version of this calculation is completely feasible. This will be the first
time that explicit expressions for this polynomial have been found.

\section{Discussion}

This paper explored several issues that arise when trying to construct
a loop representation for supergravity using the fact that the theory
can be cast as a gauge theory of the $GSU(2)$ group. There are many
detailed results that are yet to be derived, as the explicit form of
the left supersymmetry constraint in the representation constructed, a
complete set of Mandelstam identities and a suitable regularization
for the constraint. Yet, we are already able to see the emergence of a
rich mathematical structure of the representation to be constructed,
in particular concerning the space of physical states of the
theory. It is remarkable that a gauge theory with fermions yields a
state space that only includes closed loops, contrary to what happens
in other cases \cite{GaFo,MoRo}. This may be related to the extra
symmetry present in supersymmetric theories equating bosons to
fermions.  As in the non supersymmetric case it is expected that one
could find a basis of gauge invariant states that are free of
Mandelstam constraints through the use of spin networks. In this case
they would be spin networks associated with a graded group. The
properties of such objects are yet to be explored. One could also
complete the quantization of the theory in the Euclidean sector using
rigorous measure theory, as has been done in the non supersymmetric
case. Finally, as a by product we have showed that the knot polynomial
associated with Chern-Simons theory based on a graded group is the
Dubrovnik version of the Kauffman Polinomial. This is a remarkable
result; it allows new insights into that polynomial
and opens new perspectives in the search for the conjectured
Link Polinomial \cite{onknots2} which has the HOMFLY and the Kauffman
Polinomials as particular cases.

\acknowledgements

We wish to thank Luis Urrutia and Leonardo Setaro for discussions and
John Baez for pointing out several references. This work was supported
in part by grants NSF-PHY-9423950, NSF-PHY-9396246, NSF-INT-9406269,
research funds of the Pennsylvania State University, the Eberly Family
research fund at PSU and PSU's Office for Minority Faculty
development. JP acknowledges support of the Alfred P. Sloan foundation
through a fellowship. We acknowledge support of Conicyt (Uruguay) and
Conacyt (Mexico), through grant 4862-E9406.

\end{document}